\documentclass{article}

\usepackage{arxiv}

\usepackage[utf8]{inputenc} % allow utf-8 input
\usepackage[T1]{fontenc}    % use 8-bit T1 fonts
\usepackage{hyperref}       % hyperlinks
\usepackage{url}            % simple URL typesetting
\usepackage{booktabs}       % professional-quality tables
\usepackage{amsfonts}       % blackboard math symbols
\usepackage{nicefrac}       % compact symbols for 1/2, etc.
\usepackage{microtype}      % microtypography
\usepackage{lipsum}

\usepackage{amsmath,amsfonts}
\usepackage{graphicx}
\usepackage{textcomp}
\usepackage{xcolor}
\usepackage{algorithm,algpseudocode}
\usepackage{tikz}
\usepackage{tabularx} 
\usepackage{multirow}
\usepackage{array}
\usepackage{microtype}
\usepackage{subcaption}
\usepackage{enumitem}

\title{Coordinated Management of DVFS and Cache Partitioning under QoS Constraints to Save Energy in Multi-Core Systems}

\author{Mehrzad Nejat, Madhavan Manivannan, Miquel Peric{\`a}s, Per Stenstr{\"o}m \\ \\
\texttt{nejatm@chalmers.se, madhavan@chalmers.se, miquelp@chalmers.se, per.stenstrom@chalmers.se}\\ \\
Department of Computer Science \& Engineering, Chalmers University of Technology, Gothenburg, Sweden}

\newcommand{\submittedto}{Submitted to: Journal of Parallel and Distributed Computing (Nov 2019)}

\begin{document}
\maketitle

\begin{abstract}
Reducing the energy expended to carry out a computational task is important. In this work, we explore the prospects of meeting Quality-of-Service requirements of tasks on a multi-core system while adjusting resources to expend a minimum of energy. This paper considers, for the first time,  a QoS-driven \textbf{coordinated} resource management algorithm (RMA) that dynamically adjusts the size of the per-core last-level cache partitions and the per-core voltage-frequency settings to save energy while respecting QoS requirements of every application in multi-programmed workloads run on multi-core systems. It does so by doing configuration-space exploration across the spectrum of LLC partition sizes and Dynamic Voltage Frequency Scaling (DVFS) settings at runtime at negligible overhead.  We show that the energy of 4-core and 8-core systems can be reduced by up to 18\% and 14\%, respectively, compared to a baseline with even distribution of cache resources and a fixed mid-range core voltage-frequency setting.  The energy savings can potentially reach 29\% if the QoS targets are relaxed to 40\% longer execution time.
\end{abstract}

% keywords can be removed

\section{Introduction} \label{sec_intro}

Resource management, at the micro-architectural level, aims at maximizing multi-core system performance or energy efficiency. However, if applications are not associated with any  Quality-of-Service (QoS) targets, in terms of performance constraints, the energy expenditure can be excessive. In contrast, if applications have clearly defined QoS targets, resources can be throttled down to deliver enough performance with a greatly reduced energy cost.

Core Voltage-Frequency (VF) and the per-core share of the Last-Level Cache (LLC) are two popular resources to control both performance and energy-efficiency of applications running on a multi-core system. The reason is that the former is most effective for compute-intensive application phases whereas the latter can be more valuable for memory-intensive phases.

We envision a resource management system where all applications in a multi-programmed workload on a multicore system have QoS constraints that can be met by a baseline allocation of resources; e.g., partitioning of LLC resources evenly across cores at a given VF setting. The objective of our envisioned resource manager is to maximize energy efficiency by dynamically distributing resources at run-time across cores.

The literature describes several Dynamic Voltage Frequency Scaling (DVFS) and LLC partitioning resource-management schemes. Many of these schemes, such as \cite{Qureshi2006Utility-BasedCaches,Dybdahl2007AnMultiprocessors,Bitirgen2008CoordinatedApproach,Wang2015XChange:Architectures,Wang2016ReBudget:Reassignment,Jain2017CooperativeNetwork}, do not consider QoS constraints. 
Other approaches focus on optimizing the system when only a \textbf{single} application has QoS constraints \cite{Choi2002Frame-basedDecoder,Hughes2001SavingApplications,Suh2015DynamicProcessors,Chen2010MemoryOptimization,Pothukuchi2016UsingArchitectures} and do not consider QoS targets of multiple applications on a multi-core system.
A common scenario for QoS-constrained workloads is to share the system between one latency critical job and other best-effort batch jobs \cite{Iyer2007QoSPlatforms, Chen2011PredictiveMultiprocessors, Manikantan2012ProbabilisticPriSM, Funaro2016Ginseng:Allocation, Moreto2009FlexDCP:Architectures, Kasture2014Ubik, Kasture2015Rubik:Systems, Lo2015Heracles:Scale, Rahmani2018SPECTR:Management, Junokas2018CALOREE:Energy}. In this scenario, the optimization is performed on the best-effort jobs by utilizing the system resources only when the latency-critical job is not using them. In contrast, in this work, we consider the more general and challenging scenario in which all the applications in a workload have strict performance constraints. Instead of maximizing the aggregate performance or system utilization, our target is to minimize the system energy without sacrificing the performance of any application. The solution to this problem also works for a less strict scenario where a bounded reduction in performance can be tolerated on any subset of the applications.  Hence, it is a general approach.

There are only a few prior works that impose performance constraints on all the applications in a multi-programmed workload \cite{Takagi2009CooperativeMultiprocessors, Fu2011Cache-AwareSystems}. However, in these works, the core DVFS controller targets performance constraints, while the LLC controller attempts to minimize the overall number of cache misses independently from the DVFS decisions. Since minimizing the global LLC miss rate can be in conflict with meeting individual performance targets, these approaches can potentially lead to QoS violations and are thus not acceptable solutions. Furthermore, they cannot optimize system energy because the partitioning controller does not take the DVFS effect into account. For example, a smaller allocation of cache to an application can result in an increase in the core voltage-frequency which has a quadratic effect on core energy consumption.

This paper proposes, for the first time, an integrated resource manager that controls LLC partitioning and core DVFS of all the applications sharing multicore system resources in a coordinated fashion. The goal is to minimize system energy while meeting performance targets of every application. To this end, the Resource Management Algorithm (RMA) performs a configuration-space exploration, at regular program intervals, to identify the best allocation of resources. The challenge is to perform this search in a complex multi-dimensional configuration space with negligible run-time overhead and in a scalable fashion. We propose a multi-layer pruning heuristic to perform this operation in polynomial time. The proposed method does not require any profiling, training, or prior knowledge about the run-time behavior of applications.

Our experimental results show that the proposed scheme using combined DVFS and LLC partitioning is more effective in saving energy than using isolated DVFS or cache partitioning. The energy of 4-core and 8-core systems is reduced by up to 18\% and 14\%, respectively, when the QoS target is set to the baseline execution for all applications. Furthermore,  the energy savings can potentially improve up to 29\% if the QoS targets are relaxed to 40\% longer execution time.

This work makes the following contributions: 
\begin{enumerate}
    \item An online resource management scheme that controls per-core DVFS settings and LLC partitioning in a coordinated fashion to maximize system-level energy-efficiency while respecting the QoS constraints for all applications in a multiprogrammed workload.
    \item A heuristic algorithm to find an optimal resource setting in polynomial time that allows a large number of configurations to be assessed at low overhead.
    \item Evaluation of the resource management scheme, via a novel simulation framework, that compares its efficiency with different resource management algorithms on full executions of benchmark applications in a multi-core system. It provides insights on how the proposed scheme can can achieve significant energy savings.
\end{enumerate}

%This paper is an extended version of Nejat et al. \cite{Nejat2019}. This paper extends the original paper in several ways. First, we study the trade-off between QoS targets and energy savings for a wide range of workloads according to application characteristics.Second, we extend this study to scenarios where the QoS target is relaxed only for a subset of workload. Finally, we also analyze the effect of different baseline settings on potential energy savings.

This paper is an extended version of Nejat et al.~\cite{Nejat2019}. The original paper is extended in several ways. First, we perform a statistical analysis on the probability and expected value of QoS violations due to modeling error both at short-term intervals and full execution of benchmarks. Next, we study the trade-off between QoS targets and energy savings for a wide range of workloads according to application characteristics. We show that a limited reduction in the performance target (around 40\% longer execution time) significantly improves energy savings. However, we also observe that energy savings quickly saturate after this point. %But, the improvements reach a saturating point after that in many cases. 
Second, we extend this study to scenarios where the QoS target is relaxed only for a subset of workloads. We show that energy savings in a half-relaxed workload is usually near the mid-range between fully-strict and fully-relaxed workloads. %, using the proposed RMA. 
This would allow a service provider to trade-off energy cost and QoS for each user in a predictable manner. Furthermore, we analyze, as a function of different RMAs, what is the best choice of applications as victims to relax their QoS targets in order to maximize energy savings. According to the experimental results, with a RMA that controls only DVFS it is usually most effective to select memory-intensive applications. On other hand, if only cache partitioning is applied, selecting compute-intensive applications as victims in several workloads leads to the same result as selecting all applications. Using the proposed combined RMA, the energy savings improves considerably when selecting either application category as victims. The energy saving with this RMA does not dependend as much on the choice of victim applications as the other two in many of the workloads and it is in the mid-range between all-strict and all-relaxed QoS targets.
Finally, we also analyze the impact of different baseline VF settings on potential energy savings. According to our evaluations, if the performance target is increased by selecting a higher VF setting as the baseline, the proposed RMA can save a larger percentage of system energy in a majority of cases.

The rest of the paper is organized as follows. Section \ref{sec_motivation} provides the motivation for this work. The proposed scheme is described in Section \ref{sec_proposed_scheme}. The simulation methodology and experimental results are presented and discussed in Sections \ref{sec_exper_method} and \ref{sec_eval}, respectively. Then, Section \ref{sec_related_work} puts this work in perspective of related work.  Finally, we conclude in Section \ref{sec_conclude}.

\section{Motivation} \label{sec_motivation}

This section first presents the baseline architecture framework and the basic assumptions. It then provides motivational data for the potential of saving energy when both LLC and DVFS are managed in a coordinated fashion.

We consider a multi-core system where each core runs a single-threaded program. In the baseline system, the LLC capacity is evenly partitioned among the cores and all the cores run at some fixed base frequency. The QoS target for each application is expressed in terms of the instruction per second (IPS) rate on the baseline resource setting in each execution interval. This is described in detail in Section \ref{subsec_perf_model}. A cloud provider could, for example, sell computing cycles cheaper if, say, the QoS target can be relaxed to 40\% longer execution time compared to the baseline resource-setting. This form of QoS requirement can support mixed workloads with different performance targets. The RMA attempts to dynamically select a resource setting, in terms of a per-core LLC partition size and VF for each individual application, that minimizes the system energy and yet meet the QoS targets expressed as a performance constraint.

The hypothesis is that the energy savings of an isolated DVFS or LLC partitioning strategy are limited, and that with a global and coordinated control of both resources it becomes possible to find a more efficient set of resource settings. To investigate this, we conduct an experiment on different 4-core workload mixes. The details of the simulation methodology are provided in Section~\ref{sec_exper_method}. An application belongs to one of four categories: 

\begin{enumerate}[label=\Alph*-,leftmargin=2\parindent]
    \item Memory Intensive \& Cache Sensitive
    \item Memory Intensive \& Cache Insensitive
    \item Compute Intensive \& Cache Sensitive
    \item Compute Intensive \& Cache Insensitive
\end{enumerate}

An application is \textit{Memory Intensive} if it has an MPKI with the base LLC partition size that exceeds a preset threshold. Otherwise, it is counted as \textit{Compute Intensive}. In addition, an application is \textit{Cache Sensitive} if the variation in MPKI when changing from a smaller partition size to a larger one exceeds another preset threshold, relative to the per-core LLC size of the baseline system. Otherwise, it is \textit{Cache Insensitive}. The  thresholds are defined in Section~\ref{sec_exper_method}. 

Three RMAs are considered in this experiment: 1) DVFS only, 2) LLC partitioning only, and 3) Combined, i.e.,  control of DVFS and LLC partitioning in a coordinated fashion -- the main contribution of this work. All three RMAs use an ideal model of performance and energy predictions of the configuration space to select a resource setting that minimizes energy while meeting the QoS targets.

\begin{figure}[!t] 
\centering
    \includegraphics[width=0.6\columnwidth]{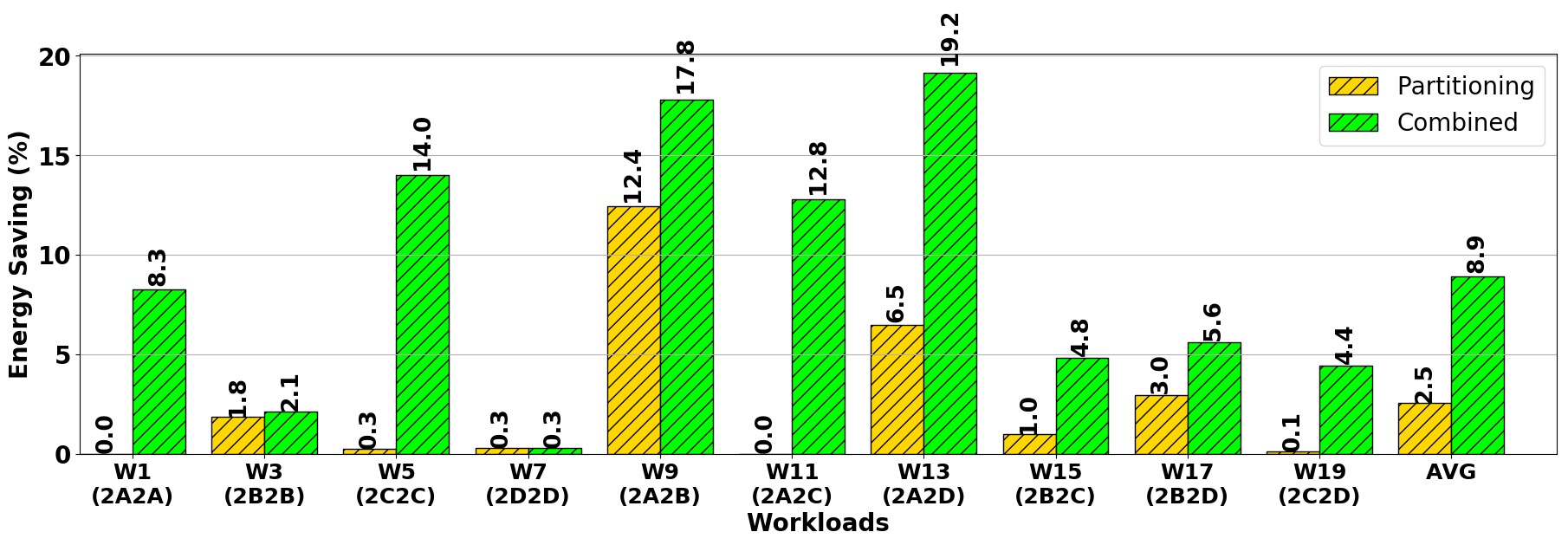}
    \includegraphics[width=0.6\columnwidth]{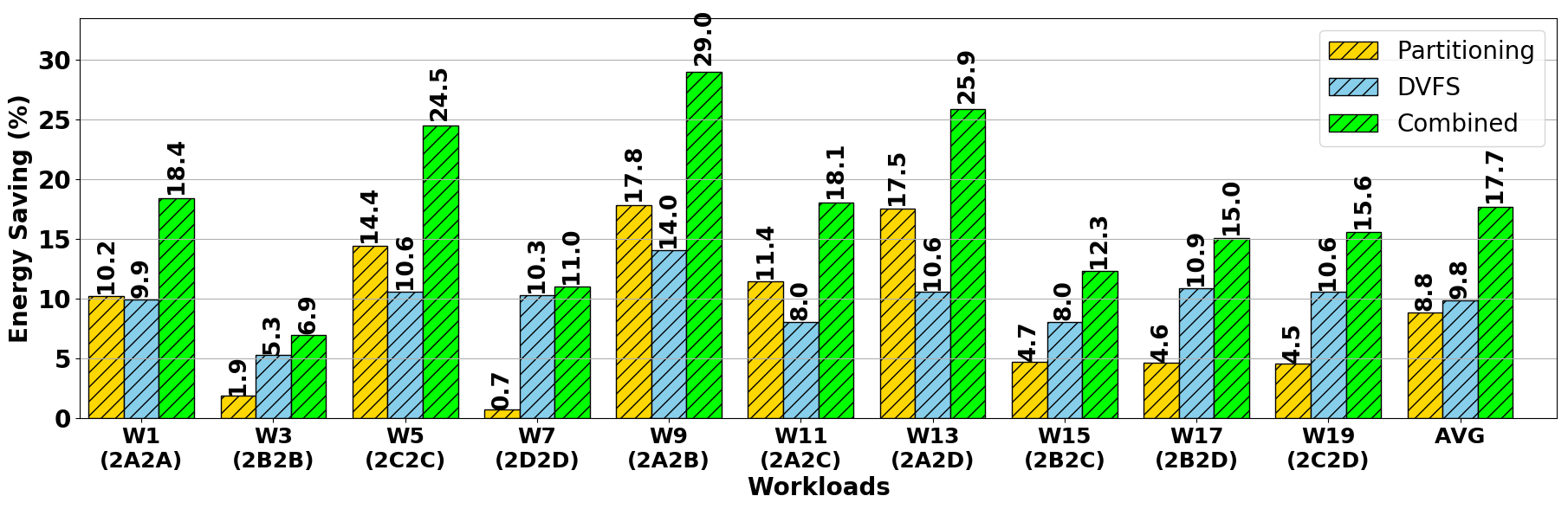}
    \caption{Energy-saving results with strict (top) and relaxed to 40\% longer execution time (bottom) QoS targets using different RMAs with perfect performance and energy prediction models neglecting the overheads.}
    \label{fig_motivational2}
\end{figure}

Figure~\ref{fig_motivational2} shows the energy saving results compared to the baseline with a strict QoS target (top) and a target relaxed to 40\% longer execution time (bottom). For each mix of application categories,  4-core workloads are randomly generated according to a methodology described in Section~\ref{subsec_workloads}. Here, we present the results for the first workload in each mix. The detailed evaluation of the complete 4-core and 8-core workload set is presented in Section~\ref{sec_eval}. 

The top figure shows that an energy saving of more than 19\% (on average 8.9\%) is possible without a performance degradation on any of the cores. Using cache partitioning alone offers only an energy saving of 2.5\%, on average. Of course, the DVFS controller cannot save energy without lowering the performance.  The combined scheme can cancel the performance degradation on the core with a smaller cache share by increasing its VF while the performance boost on the core that receives a larger cache allows a reduction of VF on that core to save energy. Thus, a more efficient resource setting that reduces the sum of core and memory access energies with the same level of performance can be found.

When all the applications are cache insensitive (B or D), none of them benefits from a larger cache and there is limited opportunity to save energy. The bottom diagram of Figure~\ref{fig_motivational2} shows that a relaxation of the QoS targets, corresponding to 40\% longer execution time,  opens up for further possibilities to save up to 29\% of energy with the coordinated scheme proposed in this work (17.7\%, on average).

\section{The Proposed Scheme} \label{sec_proposed_scheme}
\begin{figure}[t]
\centering
\includegraphics[width=0.6\linewidth]{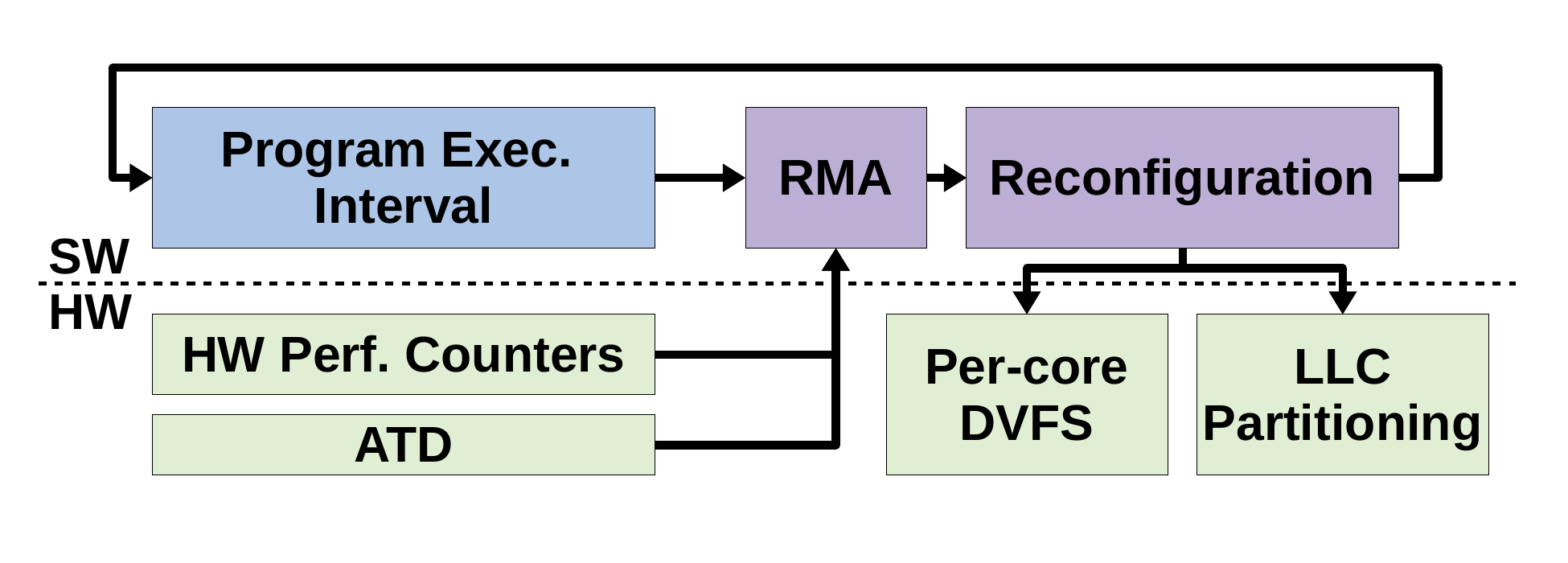} \caption{Overview of the resource management scheme.}
\label{fig_sys_overview}
\end{figure}

This section presents the proposed resource management scheme. Figure~\ref{fig_sys_overview} shows an overview of the system. On each core, a monitoring mechanism periodically collects information from hardware performance counters. The RMA, which is part of a light-weight power management software handler, is invoked at regular intervals after executing a fixed number of instructions. It uses data collected from performance counters and Auxiliary Tag Directories (ATD) \cite{Qureshi2006Utility-BasedCaches} to do configuration-space exploration of the performance and energy across all different LLC and frequency configurations. Once the new optimal configuration is found, it is applied to the DVFS and LLC partitioning controllers. The rest of this section reviews the required hardware support and the necessary software components including the SW integration, performance and energy models, and the RMA.

\subsection{Hardware Support and Software Integration} \label{subsec_hw_support}

In order to support per-core DVFS, we assume that the chip has as many voltage regulators as the number of cores. This has been implemented for example in \cite{Gupta2008SystemRegulators,Jevtic2015Per-CoreProcessors,Bowhill20154.5Family}. 
The proposed scheme requires hardware support for partitioning the LLC and predicting the miss counts at different allocations with minimum runtime overhead. We assume a partitioning of LLC ways that is for example implemented in Intel \cite{Herdrich2016CacheFamily} and Qualcomm \cite{QC2400} products. This technique has two advantages. First, the overhead of changing the partitions is limited to re-writing a bit-mask while the actual data movement is automatically performed by the replacement policy during execution. Second, it allows the use of ATDs \cite{Qureshi2006Utility-BasedCaches} to predict the effect LLC partitioning decisions on the number of cache misses. ATDs operate in parallel with the main cache and emulates the behavior of the tag directory. It predicts the number of cache misses for allocation of \textit{w} cache ways by accumulating the number of hits in recency positions larger than \textit{w} and the ATD misses. 

Furthermore, statistics from performance counters, including computation time, memory access time, number of executed instructions, and number of memory write backs, are needed. Our technique also assumes statistics to model the effect of memory-level parallelism on performance as further described in Section \ref{subsec_perf_model}. 
Finally, for implementing a low-overhead energy-model, as discussed in Section \ref{subsec_energy_model}, the system must support measurement of core energy-consumption during an execution interval.

The proposed method is invoked at the granularity of intervals with a fixed instruction count. This granularity is set to a large enough value that allows this method to be implemented as a light-weight power-management software-handler with negligible overhead. This manager's operation consists of two parts. In the first step, it collects the performance statistics by reading the registers that captures the performance counter values. In the second step, based on these statistics, it determines the VF of each core and controls the LLC partitioning by writing to the corresponding registers for allocation bit masks.

\subsection{Performance Model} \label{subsec_perf_model}
In order to predict the impact of resource allocations on the performance, we consider the following simple IPS model as a function of LLC allocation $w$ and core frequency $f$:

\begin{equation} \label{eq_ips}
    \text{IPS}(w,f)=\frac{\text{IC}}{C_{\text{base}}/f+\text{AMAT} \times M(w)}
\end{equation}

where IC is the instruction count over each execution interval, $C_{\text{base}}$ is the active CPU cycles excluding the memory access stalls, AMAT is the average memory access time of LLC misses, and $M(w)$ is the LLC miss count as a function of $w$. $C_{\text{base}}$ is derived from performance counters and we assume that it is independent of $w$. Specifically, one performance counter can capture the total stall cycles for accessing the main memory and another the total execution cycles. $C_{\text{base}}$ can be established by subtracting stall cycles from the full execution cycles. Finally, $M(w)$ is derived from the ATD.

It is well known that AMAT is sensitive to Memory Level Parallelism (MLP). To model the MLP effect, we use the approach proposed by Karkhanis and Smith \cite{Karkhanis2004AModel} based on probability functions. If $P_{\text{ov}}(i)$ denotes the probability of having $i$ overlapping LLC misses during an interval and ML is the memory access latency for an isolated DRAM access, AMAT can be calculated as follows:

\begin{equation}
    \text{AMAT}=\text{ML}\times \sum_{i} \frac{P_{\text{ov}}(i)}{i}
\end{equation}

We use this formula in our simulations by collecting the MLP histogram statistics during an interval. This can be captured by performance counters similar to those available in some modern processors (such as Intel's {\tt L1D\_PEND\_MISS.PENDING} counter). We then substitute this AMAT value in (\ref{eq_ips}) to estimate the performance of different configurations.
In Section \ref{sec_eval} we analyze the accuracy of the model.

\subsection{Energy Model}  \label{subsec_energy_model}
The RMA must only model the energy consumption of the components that are affected by its decisions. That includes the energy of core and memory accesses. This low-overhead energy-model uses the statistics collected over an execution interval with fixed number of instructions (IC). Hence the Energy-Per-Instruction (EPI) is calculated for each core $i$ as follows: 

\begin{equation} \label{eq_epi}
    \text{EPI}_i(w,f)=\frac{E_{c,\text{dyn}}(f)+P_{c,\text{static}}(f)\times T+E_{\text{mem}}(w)}{\text{IC}}
\end{equation}

In this model, $T$ is the time to execute IC instructions. This is derived from the performance model. $E_{c,\text{dyn}}$ represents the dynamic energy consumed by different core events. In our configuration space, this parameter is only affected by the core voltage which is determined by the core frequency. $P_{c,\text{static}}$ is the constant static power consumption of the core which is also dependent on the core voltage. The value of the core static power can be evaluated offline for each frequency setting and get stored in a table with as many entries as the number of frequencies. Core dynamic energy is derived by subtracting the static energy during an interval from the core energy measurements of that interval. To estimate the dynamic energy at other frequencies, this value is scaled by the core voltage squared. $E_{\text{mem}}$ is the energy consumed by memory accesses. This parameter is dependent on both the number of cache misses and write-backs to the main memory. The cache misses are estimated from the ATD, and the write-backs are measured from the performance counters. We make the simplifying assumption that the number of write-backs does not change with cache size. The accuracy of the model is studied in Section~\ref{sec_eval}

\subsection{Resource Management Algorithm (RMA)} \label{subsec_rma}

\begin{figure*}[t!]
    \centering
    \begin{subfigure}{0.6\textwidth}
        \includegraphics[width=\textwidth]{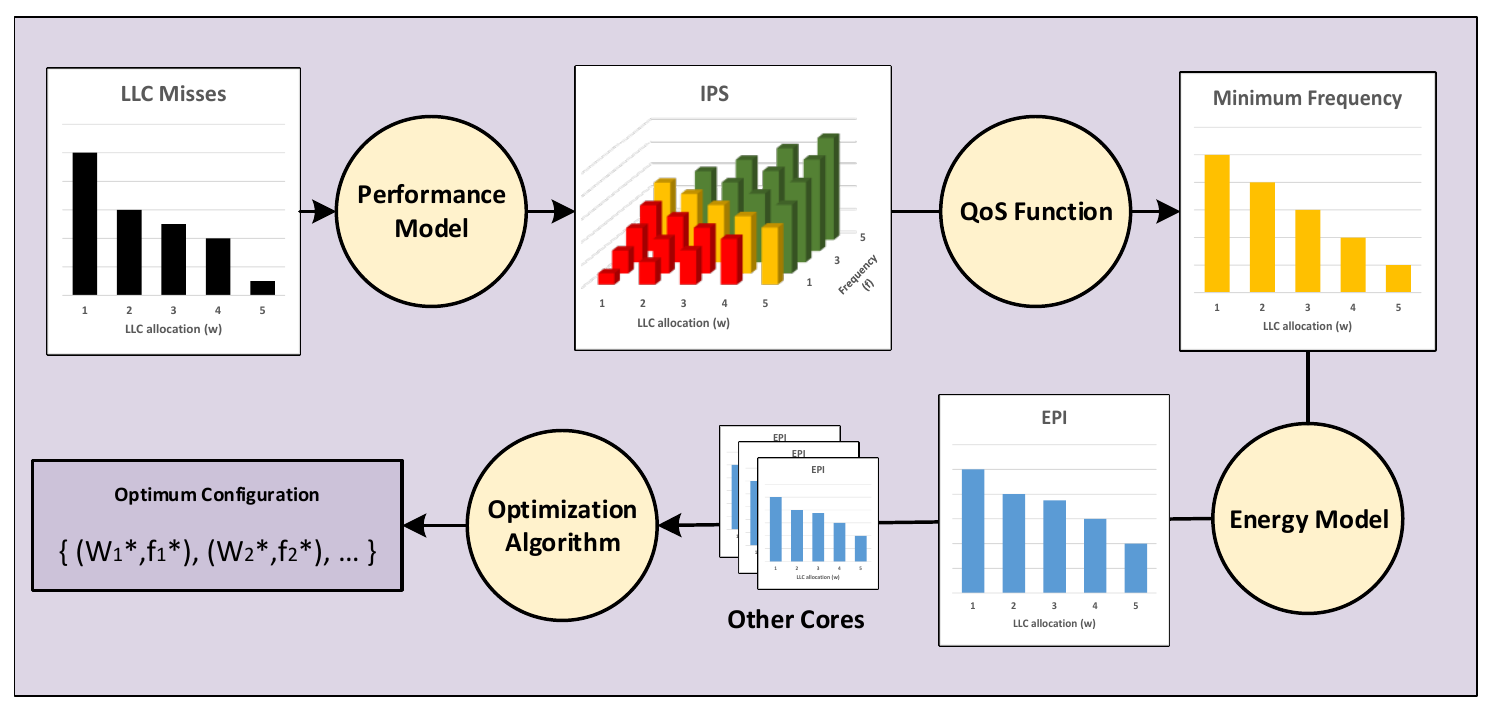}
        \caption{} \label{fig_rma_overview}
    \end{subfigure}
    \hspace{0.05\textwidth}
    \begin{subfigure}{0.2\textwidth}
        \includegraphics[width=\textwidth]{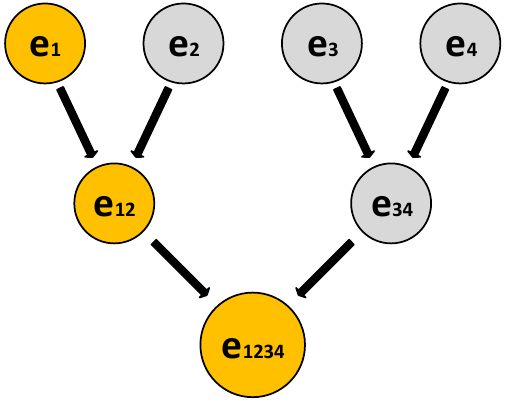}
        \caption{} \label{fig_reduction_levels}
    \end{subfigure}
    \caption{Overview of the RMA (a) and the reduction levels in the optimization algorithm for a 4 core system (b).}
\end{figure*}

An overview of the proposed RMA is shown in Figure~\ref{fig_rma_overview}. After the RMA is invoked by a particular core, LLC miss values are collected from the ATD as a function of the cache partition size. The performance model uses these values to predict the IPS rate for different system configurations. The configuration space for each core has two dimensions: cache allocation $w$ and frequency $f$. Considering all the possible combinations among all the cores creates a significantly complex search space that is not feasible for online resource management because of the performance overhead it would impose.

To address this problem, the RMA prunes the search space on each core and reduces it to a single dimension as follows. For each possible allocation of cache to each core, a minimum frequency can be found that meets the IPS constraint. This is depicted by the yellow bars in the hypothetical graphs in Figure~\ref{fig_rma_overview}. We can easily ignore the other configurations because any lower frequency violates the performance constraints while higher frequencies (and voltages) increase the energy consumption.

If $w_{b}$ and $f_{b}$ represent the baseline system configuration, the minimum frequency is derived from the following equations:

\begin{equation}
    \label{eq_qos}
    \text{QoS}(w,f)= \begin{cases}
        \text{True}, & \text{if $\text{IPS}(w,f)\geq \text{IPS}(w_{b}, f_{b})\times \alpha$}\\
        \text{False} & \text{otherwise}
    \end{cases}
\end{equation}

\begin{equation}
    \label{eq_fmin}
    f_{\text{min}}(w)= \text{\text{min} \{f $\mid$ \text{QoS}(w,f)  \}}
\end{equation}

The parameter $\alpha$ in (\ref{eq_qos}) --- $ 0 \leq \alpha \leq 1$ --- is used for relaxing the performance target. In case of a strict target, $\alpha=1$, otherwise $0 \leq \alpha < 1$. If no $f_{\text{min}}$ is found for some smaller values of $w$, those values are discarded from the minimum frequency set. 

In the next step, the energy model transforms the minimum frequency set into an EPI-set for the current core ($j$) using (\ref{eq_epi}) and (\ref{eq_fmin}):

\begin{equation}
    e_j(w_j)=\text{EPI}_j(w_j, f_{\text{min}}(w_j))
\end{equation}

At this point, the new EPI-set is passed to the optimization algorithm that already contains the EPI-sets of other cores. This algorithm finds the new optimum setting that minimizes the sum of EPI values for all the cores.

\subsection{Optimization Algorithm} \label{subsec_final_opt}

After pruning the configuration space of each core to a set of EPI values for each possible allocation of LLC, we need to find the best combination of allocations with a sum equal to the LLC size. We define a vector $V=\{w_1, w_2, ..., w_N\}$ as an LLC allocation over $N$ cores, $A$ as the total number of available LLC ways, and $W_{max}$ as an upper bound for LLC allocation to each core. We then define the optimization problem as follows:

\begin{equation}
    \begin{aligned}
        & \underset{V}{\text{minimize}} & & \sum_{j=1}^{N} e_j(w_j) & & \\
        & \text{subject to} & & \sum_{j=1}^{N} w_j = A \text{ , }& & \\
        &  & & 2 \leq w_j\leq W_{\text{max}} & \forall j \in [1,N]
    \end{aligned}
\end{equation}

Here, we assume a minimum allocation of two cache ways for each core. To solve this problem in polynomial time, we leverage the idea presented in \cite{Funaro2016Ginseng:Allocation} to design our optimization algorithm. The pseudo code is shown in Algorithm~\ref{alg_opt}. It starts from $N$ different energy curves for each core (lines 13-16). Each pair of curves are then reduced to a single curve that gives the lowest energy for a given allocation to the pair (lines 18-20). This leads to $N/2$ remaining curves. By repeating the same process, in $\log_2 N$ levels of reduction, the minimum energy configuration is found (lines 22-23).

\begin{algorithm}
    \caption{Global Optimization Algorithm Pseudocode.}
    \label{alg_opt}
    \footnotesize
    %\tiny
    \begin{algorithmic}[1]
        \State \textbf{Definitions:}
        \State $e_j(w_j)$: energy curve of core $j$ as a function of LLC way allocation $w_j$
        \State $X$: a core group e.g. \{1,2\}
        \State $V_X = \{w_j, \forall j \in X\}$ : an allocation vector to group $X$
        \State $W_X$: total allocation to group $X$
        \State $W_{\text{max}}$: maximum allocation limit to each core
        \State $T_X(W_X) = [E_X, V_X]$ : an array of tuples such that :
        \State\hspace{\algorithmicindent}$\sum\limits_{\forall w_j \in V_X} w_j=W_X$ and $\sum\limits_{\forall w_j \in V_X} e_j(w_j) = E_X$
        \State $A$: Total LLC ways, i.e. associativity
        \State $N$: Total number of cores
        \State
        \Function{Main}{  }
            \For {$j \in [1,N]$}
                \State $T_j(w_j) \gets [e_j(w_j),{w_j}]$
            \EndFor 
            \State ArrayT$\gets \{ T_j, \forall j \in [1,N]\}$
            \Repeat
                \State \textbf{For each} pair $Z=\{ j , k\}$ in ArrayT \textbf{do}
                    \State\hspace{\algorithmicindent}$T_Z\gets$ \textbf{Reduce} ($T_j,T_k$)
                    \State\hspace{\algorithmicindent}Replace $\{T_j,T_k\}$ with $T_Z$
            \Until length(ArrayT)$> 1$
            \State $T_{\text{Final}}=$ArrayT[0]
            \State \textbf{Return} $T_{\text{Final}}(A)$
        \EndFunction
        \State        
        \Function{Reduce}{$T_X,T_Y$}
            \State $n \gets \text{length}(X)+\text{length}(Y)$
            \State $W_{\text{min}}\gets n\times2$    // minimum allocation of two ways for each core
            \For {$W_{XY} \in [W_{\text{min}}, n\times W_{\text{max}}]$ } 
                \State $E_{XY}^*\gets \infty$
                \For {$W_X \in [\text{length}(X),W_{XY}-\text{length}(X)]$}
                    \State $W_Y\gets W_{XY}-W_X$
                    \State $E_X\gets T_X(W_X)(0)$
                    \State $E_Y\gets T_Y(W_Y)(0)$
                    \If{$E_X+E_Y<E_{XY}^*$}
                        \State $E_{XY}^*\gets E_X+E_Y$
                        \State $V_X\gets T_X(W_X)(1)$
                        \State $V_Y\gets T_Y(W_Y)(1)$
                        \State $T_{XY}(W_{XY})\gets [E_{XY}^*,V_X\cup V_Y] $
                    \EndIf
                \EndFor
            \EndFor
        \State \textbf{return} $T_{XY}$
        \EndFunction
    \end{algorithmic}
\end{algorithm}

The reduction process works as follows. Let us assume a pair of cores \emph{j} and \emph{k} and a maximum way allocation $W_{jk}$ to the pair while $V_{jk}=\{w_j,w_k\}$ denotes a specific allocation to these cores. A $V_{jk}^{*}$ could easily be found that minimizes $E_{jk}=e_j(w_j)+e_k(w_k)$ such that $w_j+w_k=W_{jk}$. Hence, the two curves $e_j(w_j)$ and $e_k(w_k)$ are reduced to a single curve $E_{jk}^*$ and a corresponding allocation vector $V_{jk}^{*}$ both as a function of total allocation $W_{jk}$ (lines 27-43).

One of the advantages of this algorithm is that during a reduction level, each reduction function is independent of the others. In this system, each core invokes the RMA after executing a fixed number of instructions and the energy curve is updated only for that core. Therefore, only the reductions which are affected by this update need to be executed. As depicted in Figure~\ref{fig_reduction_levels}, only $\log_2 N$ reductions are required in a system with $N$ cores. This significantly improves the overhead and scalability of the algorithm.

\section{Experimental Methodology} \label{sec_exper_method}
We evaluate the proposed RMA using a simulation method based on SimPoint analysis~\cite{Sherwood2002AutomaticallyBehavior} and Sniper~\cite{Carlson2014AnModels} plus McPAT \cite{Li2009McPATArchitectures} simulations. In Section~\ref{subsec_base_config} we present the default architecture model that is used to derive the experimental results. Then, in Section~\ref{subsec_sim_framework}, we describe the simulation framework. Finally, Section~\ref{subsec_workloads} introduces the workloads used in the simulations followed by evaluation metrics in \ref{subsec_eval_metirc}. 

\subsection{Base Configuration} \label{subsec_base_config}
In order to have a more accurate simulation, especially for studying the impact of MLP on performance, the \textit{ROB} core model in Sniper-7.2 (released 2019) \cite{sniperwebsite} is used. Table~\ref{tbl_base_config} summarizes the architectural parameters used in our simulations. The processor model is a 4-way out-of-order core. A more aggressive core would shift the workloads towards being  more memory intensive. This would make cache partitioning alone  and our proposed combined scheme more effective as it would give more headroom for trading a smaller cache partition size for a higher frequency. The baseline system consists of four cores. We will, however, present results also for eight-core systems.

\newcolumntype{b}{>{\hsize=0.6\hsize}X}
\newcolumntype{s}{>{\centering \arraybackslash \hsize=0.1\hsize}X}

\begin{table}[h]
  %\tiny
  \caption{Baseline configuration.} \label{tbl_base_config}
  \small
  \centering
  \begin{tabularx}{\columnwidth}{s|s|b|s}
    \cline{2-3}
    & Core & 4-wide, out-of-order, 128 entry reorder buffer, 64 entry reservation station, Pentium M type branch predictor, Load-Queue size of 32 and Store-Queue size of 32  &  \\
    \cline{2-3}
    & L1-I \& L1-D & 32 KB, 64 B block size, 4-way associative, LRU replacement, 10 outstanding misses, core DVFS domain & \\
    \cline{2-3}
    & L2 & Private, 256 KB, 64 B block size, 8-way associative, LRU replacement, core DVFS domain & \\
    \cline{2-3}
    & L3 & Shared, uniform access, 8-way (2 MB) per core, 64 B block size, LRU replacement policy, global DVFS domain & \\
    \cline{2-3}
    & DRAM & 100 ns base latency, 5 GB/s bandwidth per core, contention queue model & \\
    \cline{2-3}
    & DVFS & core frequency range: 1 up to 3.25 GHz, core voltage range: 0.8 up to 1.25 V, global frequency-voltage: (2GHz, 1V) & \\
    \cline{2-3}
  \end{tabularx}
\end{table}

\subsection{Simulation Framework} \label{subsec_sim_framework}

In order to reduce the simulation time, we base our simulations on the SimPoint methodology. However, to accurately model the invocations of the RMA at each interval, we need to create a version of each benchmark application that captures its phase changes. Moreover, as our simulations use workload mixes composed of multiple programs, we must create workload mixes that accurately model the phase changes of a multiprogrammed workload.

We have adopted a method based on the idea presented by Van Biesbrouck et al.~in \cite{VanBiesbrouckASimulation} as follows. A phase trace of each benchmark program is created using SimPoint analysis. The phase trace consists of the sequences of phases that a program will visit, given that the program execution is divided into instruction sequences, denoted intervals, of a fixed length. We make the simplifying assumption that the program behavior in all the intervals of a phase is exactly the same as the representative interval of that phase selected by SimPoint. Hence, the phase trace aims at mimicking the phase changes of each benchmark program.

\begin{figure}[t]
\centering
\includegraphics[width=0.6\linewidth]{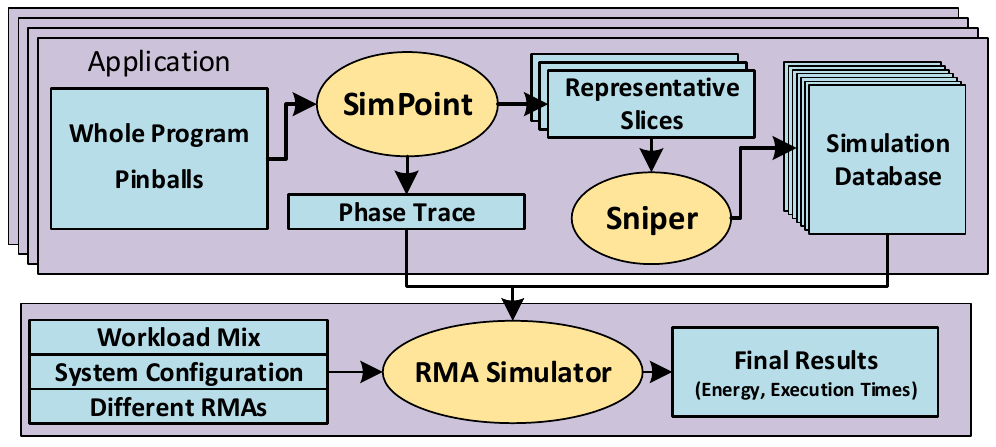} \caption{Overview of the simulation method.}
\label{fig_sim_overview}
\end{figure}

Figure~\ref{fig_sim_overview} shows an overview of the simulation steps. The SPEC CPU2006 \textit{whole program Pinballs} from the Sniper website \cite{spec_pinballs} are used as input to the process. SimPoint then generates the representative program intervals. In the next step, Sniper plus McPAT simulations are performed for representative regions of benchmark phases with 100M warm-up and 100M detailed instruction windows. These simulations are repeated over all possible VF settings, and LLC allocations (see Table~\ref{tbl_base_config} for more details). The simulation results, including detailed performance and power estimations, are collected in a database for each program phase.

\begin{figure}[t] 
\centering
\includegraphics[width=0.6\linewidth]{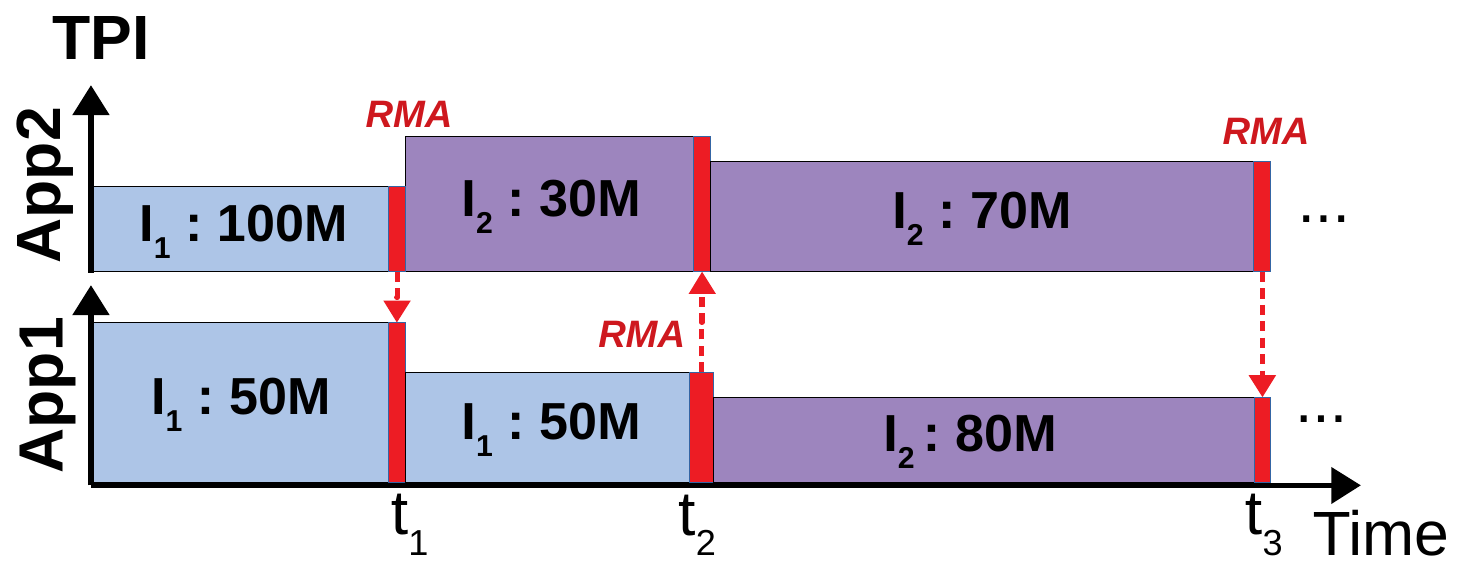} \caption{Run-time behavior of the RMA simulator.} \label{fig_sim_runtime}
\end{figure}

In the second part of the simulation process, the \textit{RMA Simulator} regenerates the execution of a multi-programmed workload in a multi-core scenario with an RMA. It uses the program-phase traces and the simulation results database for each program in the workload mix. Figure~\ref{fig_sim_runtime} shows an example scenario of a simulation run-time behavior. The simulation starts in the first program interval (I$_1$) in each application. Using the average time-per-instruction (TPI) collected from the simulation database for each application at the baseline setting, the next global event (t$_1$) is found. It is the time when the application with lowest TPI finishes one interval. The RMA is invoked on the core running that application to find a new resource setting. After updating the statistics for each core with the overhead imposed by the RMA (marked in red in Figure~\ref{fig_sim_runtime}), the next global event (t$_2$) is found in a similar fashion with the new resource settings. This process continues until the end of simulation. 

During this repetitive process, other statistics, such as energy consumption, are collected. Unlike the energy model described in Section \ref{subsec_energy_model}, the simulator collects energy consumption of the components that are shared (LLC and network-on-chip) as well as private to each core (L1 and L2 private caches), plus the dynamic energy of the main memory.
%the simulator collects both the core and un-core (LLC and network-on-chip) energy components plus the dynamic energy of the main memory.} 

\subsection{Workloads} \label{subsec_workloads}
We use \textit{SPEC CPU2006}  for our experiments.  We categorize the applications based on two important criteria: \textit{Memory Intensity} and \textit{Cache Sensitivity}. We define these criteria by considering the MPKI curve of each application for different LLC partition-sizes around the per-core baseline partition size (see Table~\ref{tbl_base_config}). We consider an application with a high base MPKI (more than 5) as memory intensive. On the other hand, to determine cache sensitivity, we measure the variation in MPKI around the baseline partition size. Specifically, if the difference in MPKI between 50\% smaller to 50\% larger LLC allocation is beyond a threshold (20\% of the baseline MPKI) while the baseline MPKI is not too small (more than 0.2), the application is counted as cache sensitive. 
%On the other hand, if the MPKI variation between 50\% and 150\% of the base is larger than a threshold (20\%) with a large enough base MPKI (more than 0.2), we count it as cache sensitive. 
Table~\ref{tbl_app_categories} shows the benchmarks that belong to each category. For two of the SPEC CPU2006 applications (\textit{calculix, milc}), the Sniper simulations did not finish properly in some phases\footnote{The simulation was aborted in the middle due an error caused by \textit{SIFT} reader.}. Therefore, they are excluded from this study.

\newcolumntype{s}{>{\centering \arraybackslash \hsize=0.1\hsize}X}
\newcolumntype{b}{>{\centering \arraybackslash \hsize=0.2\hsize}X}
\newcolumntype{m}{>{\centering \arraybackslash \hsize=0.5\hsize}X}

\begin{table}[ht]
    \small
    \caption{Application categories.} \label{tbl_app_categories}
    \centering
    \begin{tabularx}{\columnwidth}{s|s|b|m|s}
        \cline{2-4} 
        & \textbf{Type} & \textbf{Attributes} & \textbf{Benchmarks} & \\ 
        \cline{2-4}
         & A & Memory Intensive \& Cache Sensitive & mcf, omnetpp, sphinx3, xalancbmk, soplex  &  \\
        \cline{2-4}
         & B & Memory Intensive \& Cache Insensitive & leslie3d, lbm, bwaves, GemsFDTD, wrf, astar, libquantum & \\
        \cline{2-4}
        &  C & Compute Intensive \& Cache Sensitive & gobmk, gcc, h264ref, gromacs, bzip2, hmmer, tonto & \\
        \cline{2-4}
        &  D & Compute Intensive \& Cache Insensitive & dealII, namd, povray, perlbench, cactusADM, gamess, sjeng , zeusmp& \\
        \cline{2-4} 
  \end{tabularx}
\end{table}

We create a list of different combinations of application types to model a wide range of 4 and 8 core workload mixes. We then use the python function \texttt{random.choice()} to select benchmarks from each category for each workload. This process is repeated until each application is selected at least once across all the workloads. The result is listed in Table~\ref{tbl_workload_mixes}. For each 4-core mix, two workloads are randomly generated. For each 8-core mix, however, only one workload is studied due to a limited number of applications in each category as well as longer simulation time.

\newcolumntype{s}{>{\centering \arraybackslash \hsize=0.10\hsize}X}
\newcolumntype{b}{>{\centering \arraybackslash \hsize=0.8\hsize}X}

\begin{table}[h]
    \small
    \caption{Workload mixes.} \label{tbl_workload_mixes}
    \centering
    \begin{tabularx}{\columnwidth}{|s|s|b|}
        \hline
        \textbf{Label} & \textbf{Mix} & \textbf{Applications} \\
        \hline
        \hline
        \multicolumn{3}{|c|}{\textit{\textbf{4-Core}}}\\
        \hline
	    W1 & 2A2A & omnetpp, mcf, soplex, sphinx3 \\ \hline
        W2 & 2A2A & mcf, omnetpp, soplex, xalancbmk \\ \hline
        W3 & 2B2B & lbm, astar, bwaves, lbm \\ \hline
        W4 & 2B2B & libquantum, leslie3d, astar, leslie3d \\ \hline
        W5 & 2C2C & bzip2, gcc, gobmk, tonto \\ \hline
        W6 & 2C2C & h264ref, h264ref, h264ref, gobmk \\ \hline
        W7 & 2D2D & perlbench, perlbench, gamess, perlbench \\ \hline
        W8 & 2D2D & sjeng, cactusADM, dealII, gamess \\ \hline
        W9 & 2A2B & sphinx3, sphinx3, GemsFDTD, leslie3d \\ \hline
        W10 & 2A2B & soplex, xalancbmk, wrf, lbm \\ \hline
        W11 & 2A2C & xalancbmk, mcf, gobmk, gobmk \\ \hline
        W12 & 2A2C & omnetpp, xalancbmk, bzip2, hmmer \\ \hline
        W13 & 2A2D & soplex, sphinx3, zeusmp, povray \\ \hline
        W14 & 2A2D & mcf, sphinx3, zeusmp, gamess \\ \hline
        W15 & 2B2C & GemsFDTD, astar, gromacs, gromacs \\ \hline
        W16 & 2B2C & GemsFDTD, leslie3d, bzip2, gcc \\ \hline
        W17 & 2B2D & GemsFDTD, wrf, povray, cactusADM \\ \hline
        W18 & 2B2D & GemsFDTD, bwaves, perlbench, namd \\ \hline
        W19 & 2C2D & gcc, h264ref, namd, namd \\ \hline
        W20 & 2C2D & gromacs, tonto, sjeng, zeusmp \\ \hline
        \hline
        \multicolumn{3}{|c|}{\textit{\textbf{8-Core}}}\\
        \hline
        W1 & 4A4A & xalancbmk, omnetpp, mcf, omnetpp, sphinx3, xalancbmk, soplex, xalancbmk \\ \hline
        W2 & 4B4B & wrf, bwaves, bwaves, lbm, leslie3d, GemsFDTD, wrf, bwaves \\ \hline
        W3 & 4C4C & bzip2, gobmk, h264ref, gcc, gromacs, tonto, gobmk, h264ref \\ \hline
        W4 & 4D4D & sjeng, cactusADM, perlbench, namd, zeusmp, perlbench, cactusADM, namd \\ \hline
        W5 & 4A4B & omnetpp, xalancbmk, omnetpp, sphinx3, GemsFDTD, libquantum, lbm, libquantum \\ \hline
        W6 & 4A4C & mcf, sphinx3, xalancbmk, xalancbmk, bzip2, hmmer, tonto, gobmk \\ \hline
        W7 & 4A4D & xalancbmk, omnetpp, soplex, soplex, gamess, zeusmp, cactusADM, dealII \\ \hline
        W8 & 4B4C & GemsFDTD, astar, astar, libquantum, h264ref, gobmk, h264ref, gromacs \\ \hline
        W9 & 4B4D & libquantum, leslie3d, bwaves, astar, gamess, sjeng, povray, cactusADM \\ \hline
        W10 & 4C4D & hmmer, hmmer, gcc, bzip2, gamess, namd, zeusmp, gamess \\ \hline
  \end{tabularx}
\end{table}

The total number of instructions varies significantly across benchmark applications. Therefore, in order to have a fair comparison, the simulations are run until each application in the workload has executed the same number of instructions. This number is set according to the longest benchmark which consists of 4146B instructions. Once an application reaches the end of its execution, it is re-started until the end of simulation. 

\subsection{Evaluation Metrics} \label{subsec_eval_metirc}

% the following could be skipped, added to avoid an empty heading - miquelp 
This section describes the main metrics used for evaluation, namely energy savings (Sec~\ref{sec:energy_savings}) and QoS violations (Sec~\ref{subsec_metric_qos}).  

\subsubsection{Energy Saving} \label{sec:energy_savings} 
The energy consumption is calculated as the sum of the total core energy (including L1 and L2 caches) and the dynamic energy of LLC and DRAM for every application in the workload. For each application, only the energy for the execution of the predefined number (4146B) of instructions is accounted for. The static energy of the shared components (LLC and network-on-chip) is added to the results until the end of simulation for all applications. We compare against the energy of the baseline system corresponding to an idle RMA that keeps the baseline system setting until the end of simulation. The same three RMAs are evaluated as mentioned in Section~\ref{sec_motivation}. 

\subsubsection{QoS Violations} \label{subsec_metric_qos}
When the RMA is invoked at the end of each execution interval $i$, it attempts to find a target resource setting for the upcoming interval $i+1$ that satisfies QoS according to Equation~\ref{eq_qos}. This needs performance modeling of both the target setting and the baseline setting. However, due to modeling error, the RMA may select a setting that violates QoS for the next interval ($i+1$). We denote this as a \textit{short-term} QoS violation. Short-term here refers to the fact that a single interval violation is often compensated for by faster runs in other intervals and hence it will not frequently  result in QoS violations in the \textit{long-term}. 

We perform an extensive analysis to estimate the probability and expected value of short-term QoS violations over all benchmark applications as follows. For the upcoming execution interval ($i+1$), the short-term QoS is violated if the RMA selects a target resource setting (f, w) that meets two conditions: 
\begin{enumerate}
    \item The actual IPS with the target setting is smaller compared to the baseline setting ($\textit{f}_b,\textit{w}_b$):
    \begin{equation}
        \text{IPS}^{\text{Act.}}_{i+1}(\textit{f,w}) < \text{IPS}^{\text{Act.}}_{i+1}(\textit{f}_b,\textit{w}_b)    
    \end{equation}
    \item The analytical performance model predicts the IPS with the target setting to be greater or equal compared to the baseline setting: 
    \begin{equation}
        \text{IPS}_{i+1}(\textit{f,w}) = \text{IPS}^{\text{Act.}}_{i+1}(\textit{f,w}) + \text{err}_{\text{f,w}} \geq \text{IPS}^{\text{Act.}}_{i+1}(\textit{f}_b,\textit{w}_b) + \text{err}_{\text{b}}= \text{IPS}_{i+1}(\textit{f}_b,\textit{w}_b) 
    \end{equation}
\end{enumerate}

Here we assume a strict QoS target with $\alpha = 1.0$ (see Equation~\ref{eq_qos}). The modeling of interval $i+1$ is performed using the statistics collected at interval $i$. The probability of QoS violation is evaluated by iterating over all phases of all applications, all possible current settings (during interval $i$), and all possible target settings (for interval $i+1$) and checking the above conditions. The phase weights generated by SimPoint are used as the probability of each program phase. Within each phase, all current and target settings are assumed to be equally probable. In case of a short-term QoS violation, the amount of violation is calculated as follows: 

\begin{equation} \label{eq_violation}
    \text{Violation} = \frac{\text{T}^{\text{Act.}}(\text{f,w}) - \text{T}^{\text{Act.}}(\textit{f}_b,\textit{w}_b)} {\text{T}^{\text{Act.}}(\textit{f}_b,\textit{w}_b)}
\end{equation}

Here, $\text{T}^{\text{Act.}}$ denotes the actual execution time of the interval $i+1$. These values are used to calculate the expected value and standard deviation of short-term QoS violations which are reported in Section~\ref{subsec_qos_results}. 

In addition to short-term QoS, we consider full execution times of each benchmark in each workload. As mentioned earlier, during a full simulation, benchmarks are re-started multiple times. Hence, the average execution time of each benchmark is measured. This value is compared for each RMA to an idle RMA that keeps the baseline system setting until the end of simulation. An average execution time longer than the baseline is counted as a \textit{long-term} QoS violation. The result of long-term QoS violations is reported in Section~\ref{subsec_qos_results} for all the workloads.

\section{Experimental Results} \label{sec_eval}
The experimental results are reported and discussed in this section. We first perform an experiment to evaluate the energy savings with the proposed scheme in a wide range of workloads with strict performance constraints (Section~\ref{subsec_energy_savings}). The effects of modeling errors on energy saving results are studied in Section~\ref{subsec_modeling_effect}. Modeling errors may also lead to a small QoS violation in a few cases. These QoS violations are extensively analyzed and evaluated in Section~\ref{subsec_qos_results} for all workloads. The achievable energy savings, using the proposed RMA, are limited with strict performance constraints. But, if the users can tolerate a bounded reduction in performance, energy savings can improve substantially. We perform a series of experiments to study the trade-off between relaxed QoS targets and energy savings in Sections~\ref{subsec_varyingalpha} and~\ref{subsec_mixed_qos}. In the last experiment, we evaluate the sensitivity of the proposed scheme to the baseline VF setting (Section~\ref{subsec_base_sensitivity}). Finally, the impact of overheads is reported in Section~\ref{subsec_OH_results}. 

\subsection{Energy Savings with Strict QoS Targets} \label{subsec_energy_savings}
\begin{figure}[h] 
\centering
    \includegraphics[width=0.95\columnwidth]{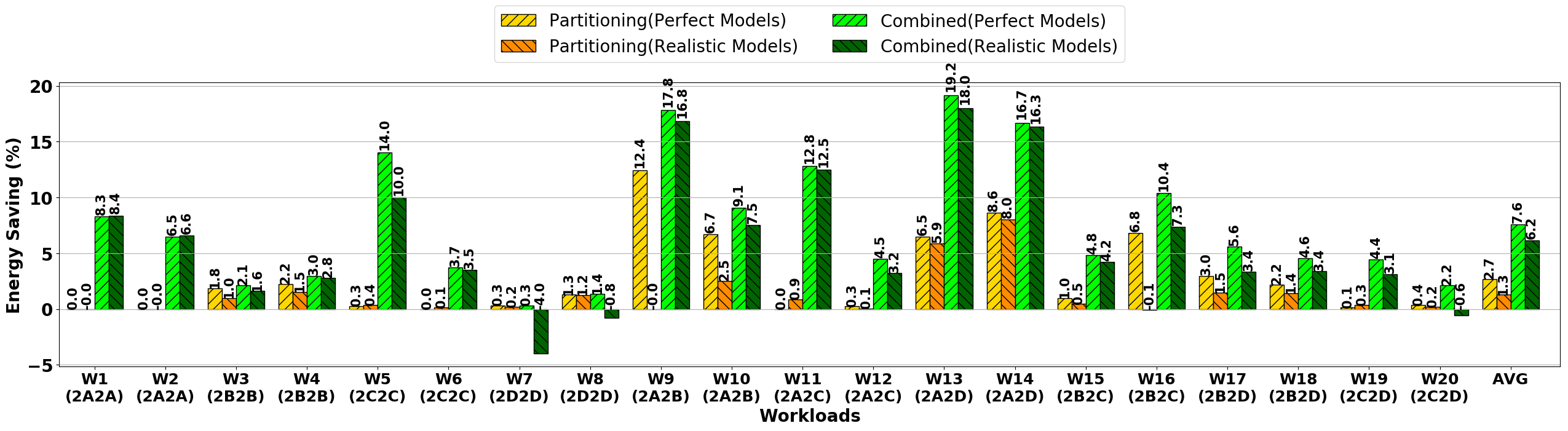}
    \includegraphics[width=0.6\columnwidth]{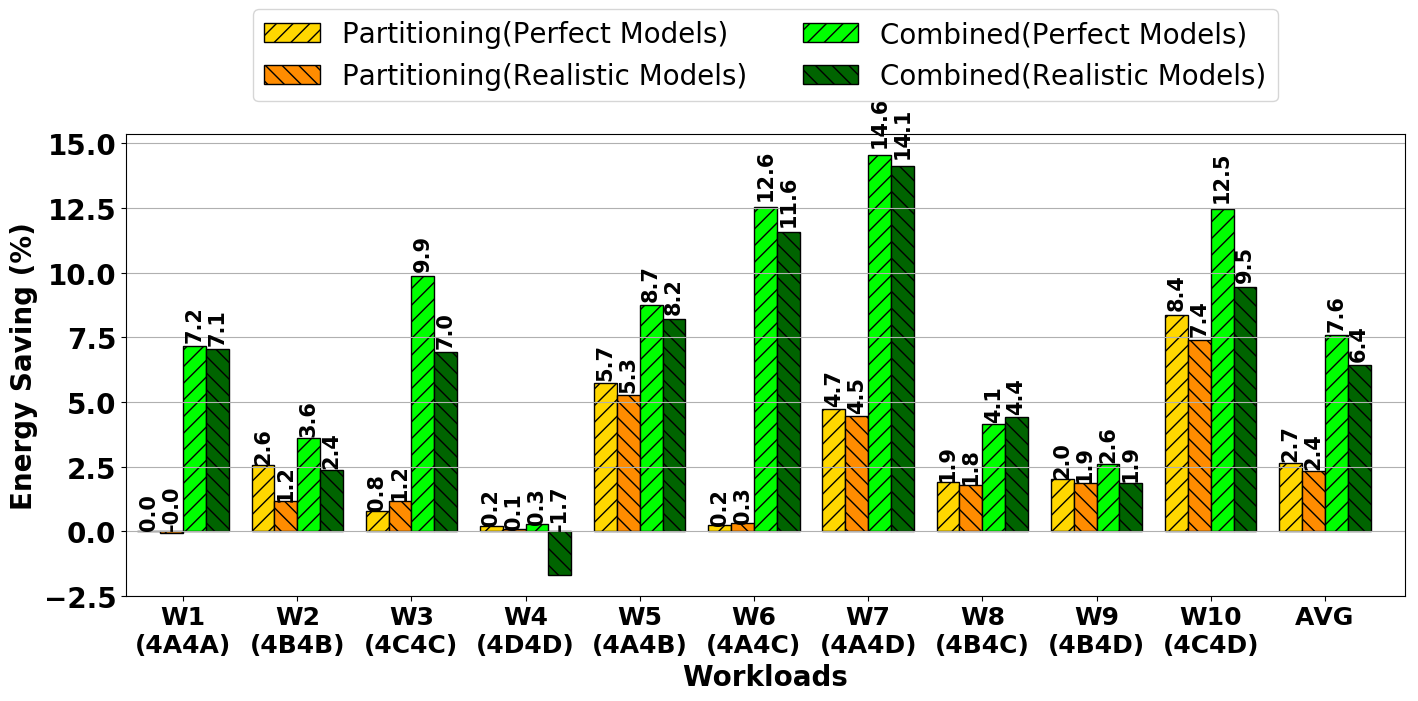}
    \caption{4-core (top) and 8-core (bottom) simulation results with strict QoS targets.}
    \label{fig_energy_strict}
\end{figure}

We perform the experiments on 4 and 8 core workloads (see Section~\ref{subsec_workloads}) with strict performance constraints; i.e., no performance degradation is allowed on any application, which corresponds to $\alpha = 1$ in Equation~\ref{eq_qos}. We consider two RMAs: i) Cache partitioning only and ii) the proposed RMA with coordinated control of DVFS and cache partitioning, called \textit{Combined}. The \textit{Partitioning} RMA controls only the LLC partitioning without affecting the core frequencies. Its goal is to minimize system energy without violating the performance constraints. It uses the same optimization algorithm described in Section~\ref{subsec_final_opt}. The DVFS-only RMA is not relevant in this scenario because it cannot affect a system with strict performance targets.

Two sets of simulations are done for each RMA: One with idealistic assumptions to show the potential with perfect performance and energy modeling and neglecting the RMA overheads and the other for a realistic system that uses the analytical performance and energy models described in Section \ref{sec_proposed_scheme} with overheads added to the simulations. The impact of overhead will be analyzed in detail later.

Figure~\ref{fig_energy_strict} shows the energy savings of the two RMAs relative to the baseline. For each workload mix, we show four bars corresponding to, from left to right, the \textit{Partitioning} with perfect models, the \textit{Partitioning} with analytical models and overheads, the \textit{Combined} with perfect models, and the \textit{Combined} with analytical models and overheads.

Overall, we can see that there is a huge potential of the \textit{Combined} scheme. Compared to \textit{Partitioning}, \textit{Combined} manages to save substantially more energy on a realistic system, with the performance and energy models proposed. On a four-core system the saving is, on average, 6.2\% versus 1.3\% whereas on an eight-core system it is 6.4\% versus 2.4\%.

We now analyze the findings in more detail. First, the energy savings offered by \textit{Partitioning} are mostly limited to workloads that are a mix of cache sensitive (A or C) and cache insensitive (B or D) applications. This is not the case for the \textit{Combined} RMA because it has a secondary dimension of flexibility; frequency variation. In the workloads that are all cache sensitive, i.e., mixes of A or C, the \textit{Combined} scheme shows a significant advantage over the \textit{Partitioning} RMA. Second, in the workloads that are all cache insensitive, i.e. mixes of B or D, none of the RMAs are very effective since any re-distribution of cache resources does neither improve the performance nor energy of any application. In fact, with limited modeling accuracy and considering the overheads, this may even lead to a small increase in the energy consumption.

\subsection{Effect of Modeling Accuracy} \label{subsec_modeling_effect}
We now evaluate the effect of modeling errors on energy savings by comparing the results with perfect models to those with realistic analytical models. As Figure~\ref{fig_energy_strict} shows, in most of the cases the modeling errors reduce the energy savings. 

In the 4-core results, this leads to a substantial reduction of energy savings with the \textit{Partitioning} RMA, especially for W9, W10, and W16. In these workloads, the cache insensitive applications (B) can give up a portion of their LLC share to other applications that benefit from it. However, the error in performance modeling prevents the RMA from exploiting this trade-off in order to meet the QoS target. The \textit{Combined} RMA, on the other hand, is not affected as much since it can search a larger configuration space with the second dimension (i.e.~DVFS) to achieve energy savings even in the presence of modeling error.

The 8-core results show a different trend. On average, the modeling error has smaller impact on \textit{Partitioning} compared to \textit{Combined}. With the larger number of applications in the workload, it is more likely that cache-sensitive applications can get a larger LLC share by finding unused LLC ways in cache-insensitive applications. Therefore, the results for \textit{Partitioning} RMA are not affected by modeling error as much as 4-core workloads. The effect of error on the \textit{Combined} RMA is on average similar in both 4-core and 8-core workloads. 

There are a few cases in which modeling errors lead to a negligible increase in energy saving (W11, and W19 in 4-core plus W3, W6, and W8 in 8-core). With a low probability, a performance modeling error can lead to QoS violation which is discussed in Section~\ref{subsec_qos_results}. In that case, the RMA may select a resource setting that further reduces system energy, at the expense of a small reduction in performance. This setting, however, is never selected if perfect models are used.

\subsection{QoS Evaluation}\label{subsec_qos_results}
As mentioned in Section~\ref{subsec_metric_qos}, we study two forms of QoS: i) \textit{long-term} that considers full execution of each benchmark during the simulation and ii) \textit{short-term} that focuses on each execution interval (100M instructions).

We first analyze the effect of modeling error on \textit{short-term} QoS. This analysis is performed according to the methodology explained in Section~\ref{subsec_metric_qos}. The analysis over all phases of all benchmark applications, shows a probability of 4.3\% for a \textit{short-term} QoS violation in the up-coming interval. Considering only the violating cases, the expected value of violation (see Eq.~\ref{eq_violation}) is 3.4\% with a standard deviation of 14.9\%. This comprehensive analysis is independent of the RMA and assumes equal probability for any target resource setting. However, during program execution, many of these settings are never selected by the RMA. Furthermore, the selected resource setting for many intervals may result in faster execution compared to the baseline. This will cancel a part of the \textit{short-term} violations in the \textit{long-term} run. 

Next, we report the \textit{long-term} QoS results for the simulations in Section~\ref{subsec_energy_savings}. With the \textit{Combined} RMA that uses the analytical models, in 13 cases out of 80 applications in the 4-core workloads, modeling errors lead to an average execution time longer than the baseline by more than 1\%. The average value of violation among these 13 cases is 3.0\% with a maximum of 9.3\%. Considering the 8-core results with the \textit{Combined} RMA, 15 violations are detected out of 80 applications. The average violation is 2.7\% with a maximum of 6.8\%. Modeling errors do not lead to a considerable \textit{long-term} violation with the \textit{Partitioning} RMA as it has low flexibility compared to \textit{Combined} RMA.

\subsection{Energy Performance Trade-off}\label{subsec_varyingalpha}
\begin{figure}[h] 
\centering
    \includegraphics[width=\columnwidth]{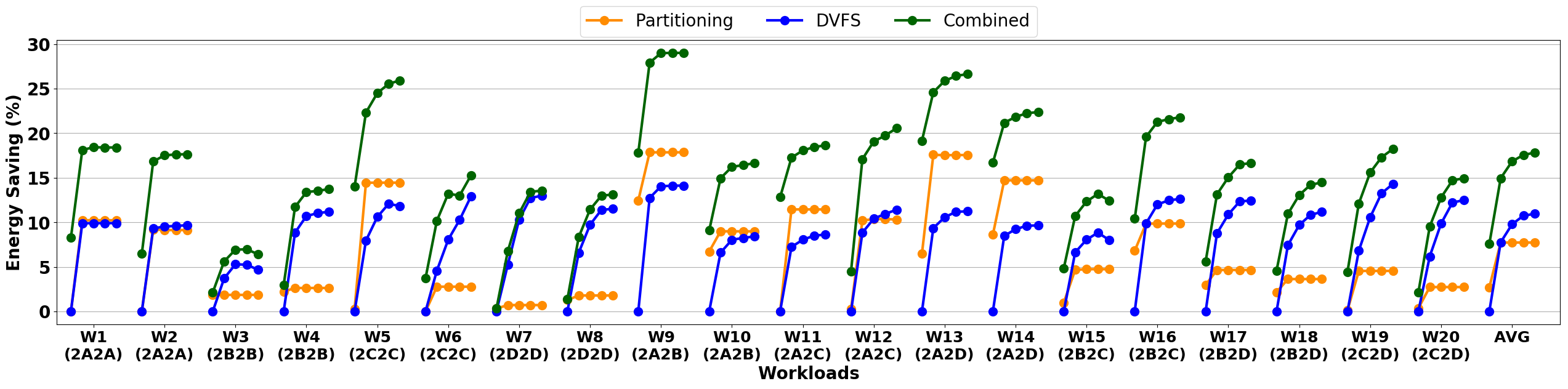}
    \caption{Energy savings for different levels of QoS relaxation.}
    \label{fig_varyingalpha}
\end{figure}

Even though there are energy-saving possibilities by optimizing the resource trade-offs between applications, the total amount of saving is limited without trading off performance. If the user can tolerate a bounded reduction of performance, further energy savings become possible. To evaluate this effect, we gradually relax the performance constraint of all applications in 4-core workloads, using the $\alpha$ parameter in Equation~\ref{eq_qos}. Figure~\ref{fig_varyingalpha} shows the resulting trends in energy savings for three different RMAs. In addition to the \textit{Partitioning} and \textit{Combined}, a RMA that performs only per-core \textit{DVFS} is also present. The \textit{DVFS} RMA can save energy if the performance constraint is relaxed. In order to study the trends in potential energy savings in the absence of modeling error and QoS violations, perfect models are used in this experiment. For each workload, three curves are presented that correspond to each RMA. The points on each curve, from left to right, represent $\alpha$ values that correspond to 0\%, 20\%, 40\%, 60\% and 80\% longer execution time, i.e.~$\alpha = \{\frac{1}{1.0},\frac{1}{1.2},\frac{1}{1.4},\frac{1}{1.6},\frac{1}{1.8}\}$. 

An important observation in this experiment is that energy savings usually saturate after a certain amount of relaxation in performance constraints. In a few cases (W3, W5, W6, and W15) there is even a small reduction in energy saving after a certain point. This is a result of larger LLC leakage energy, a component that is not accounted for in the RMA. In the case of W6, the additional energy savings when reducing $\alpha$ from $\frac{1}{1.4}$ to $\frac{1}{1.6}$ are not as high as the increase in LLC leakage. But a further decrease of $\alpha$ to $\frac{1}{1.8}$ opens a new resource trade-off that improves energy savings beyond the increase in LLC leakage. The saturation usually occurs earlier for \textit{Partitioning} compared to the other two RMAs. This shows that a limited relaxation of constraints enables the most efficient distribution of LLC resources to minimize system energy at a fixed core VF. However, \textit{DVFS}, in general, has a stronger impact on energy consumption except for workloads with memory-intensive and cache-sensitive applications (A) and one case with compute-intensive and cache-sensitive applications (W5). The \textit{Combined} RMA, outperforms the other two in all the workloads and substantially improves energy savings. It can potentially save up to 29\% of system energy with only a 40\% relaxation of the performance target (W9). On average, it can save up to 18\% of energy compared to 11\% with \textit{DVFS} and 8\% with \textit{Partitioning}. 

\subsection{Mixed QoS workloads}\label{subsec_mixed_qos}

\begin{figure}[h] 
\centering
    \includegraphics[width=\columnwidth]{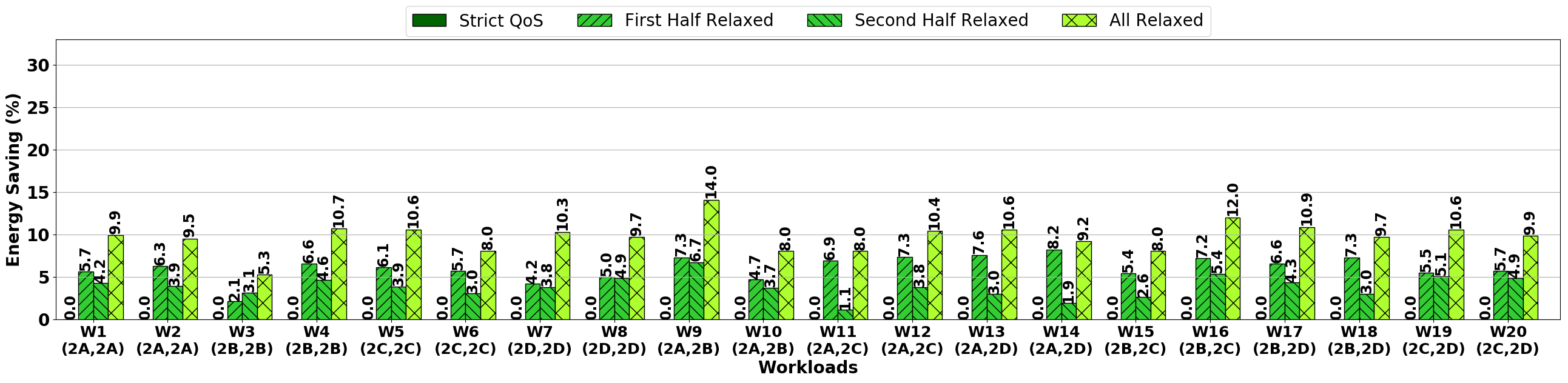}
    \includegraphics[width=\columnwidth]{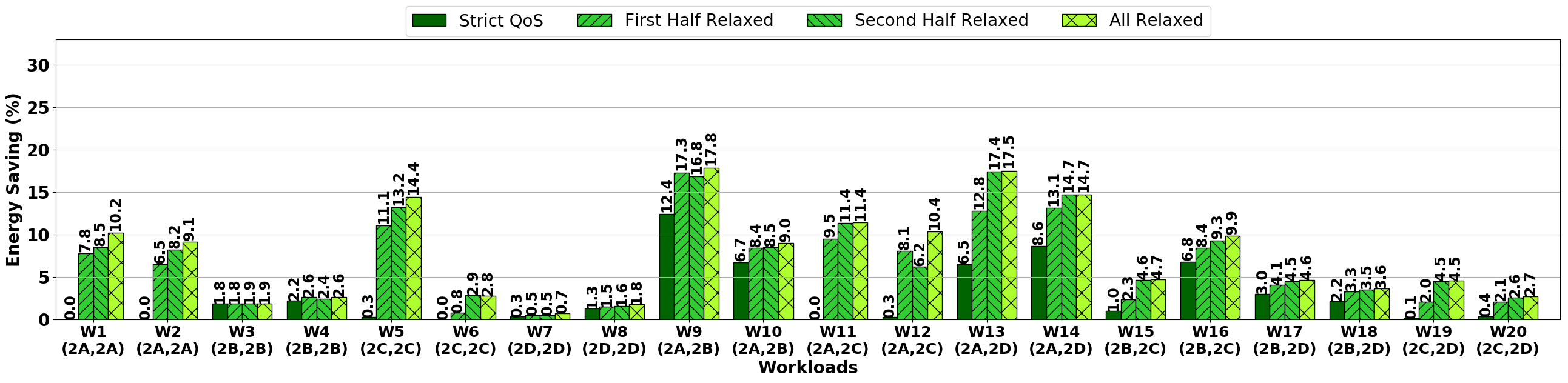}
    \includegraphics[width=\columnwidth]{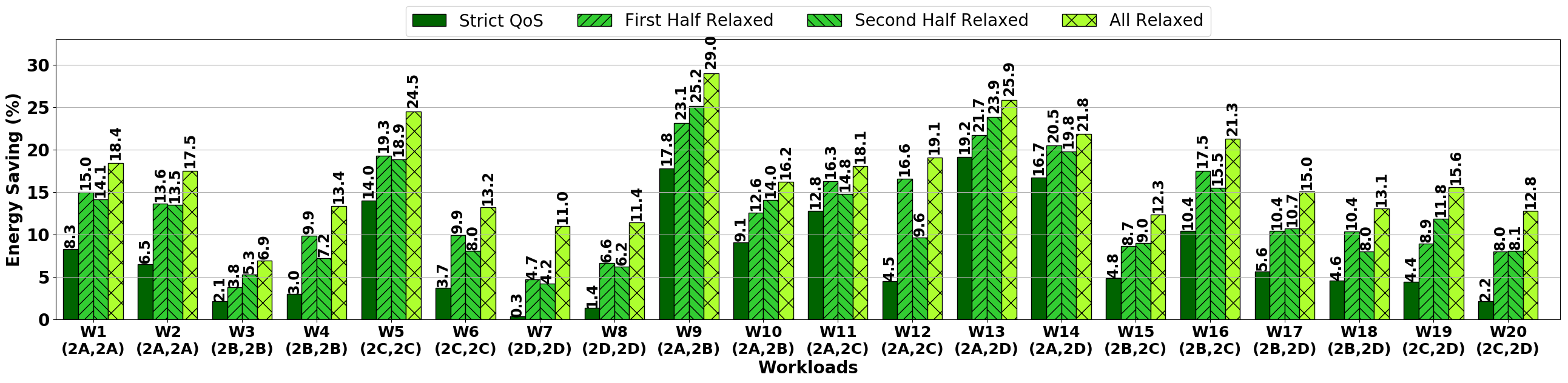}
    \caption{Energy Savings for mixed QoS scenarios with DVFS alone (top), LLC partitioning alone (middle), and combined (bottom) RMAs when using perfect models.}
    \label{fig_mixedQoS}
\end{figure}

In the previous experiment in Section~\ref{subsec_varyingalpha}, the QoS target is relaxed for all applications in each benchmark. However, it may not be possible for all the users that share the system to tolerate a performance degradation. 
Even if only a subset of the users can tolerate a bounded reduction in performance, energy savings may improve considerably. The ability to select a subset of the workload as a victim for bounded performance degradation increases the flexibility of the service provider to make trade-offs between the QoS delivered to each individual user and the system energy consumption. In that case, an important question is to find out which subset should be selected as a victim to achieve the highest energy saving. Therefore, in this experiment, we evaluate two scenarios in 4-core workloads. 

In one scenario the QoS target is relaxed only for the first half of the applications in the workload, whereas in the other scenario it is relaxed only for the second half. The results are depicted in Figure~\ref{fig_mixedQoS} for three different RMAs. Similar to Section~\ref{subsec_varyingalpha}, perfect models are used in this experiment. There are four sets of bars in each figure that correspond to, from left to right, strict targets for all applications, relaxed targets for the first half, relaxed targets for the second half, and relaxed targets for all applications. In all these cases, a relaxed target corresponds to an $\alpha$ value of $\frac{1}{1.4}$. 
This value is selected based on the observations in Section~\ref{subsec_varyingalpha}. 

According to Figure~\ref{fig_mixedQoS} (top), with the \textit{DVFS} RMA, relaxing the target for memory-intensive applications (A and B) can be more beneficial compared to compute-intensive applications (C and D). This is evident in workloads W11 to W18. Due to the reduced effect of DVFS on the performance of memory-intensive applications, a limited relaxation of the QoS targets enables a substantial reduction in core VF and energy consumption. 

On the other hand, the \textit{Partitioning} RMA (Figure~\ref{fig_mixedQoS} - middle) shows a different result. In most of the workloads that are mixes of memory-intensive and cache-sensitive applications (A) with compute-intensive applications (C and D), the highest energy saving is achieved if the target is relaxed only for the latter subset. In this case, the relaxation enables compute-intensive applications to give up their LLC share to type A applications that benefit more from it. In the case of W12, the type A applications are \textit{omnetpp} and \textit{xalancbmk}. Even though the baseline MPKI is high for these applications and the MPKI variation is above the cache sensitivity threshold (see Section \ref{subsec_workloads}), their MPKI does not reduce as much as \textit{bzip2} (one of the C applications) when receiving a larger LLC share. Therefore, in this case, it is more beneficial to relax the target for the first half of the workload. In general, with the \textit{Partitioning} RMA, if the target is relaxed for only half the workload, the energy saving is comparable to the case where all the targets are relaxed. 

Using the \textit{Combined} RMA (Figure~\ref{fig_mixedQoS} - bottom), the energy savings increase substantially in all the cases compared to the other two RMAs. This RMA has a higher level of flexibility with two different resources. Therefore, in most of the cases, both half-relaxed workload scenarios lead to comparable energy savings. In general, the energy saving in these two scenarios are near the arithmetic average between fully-strict and fully-relaxed workloads.

\subsection{Sensitivity to Baseline Setting}\label{subsec_base_sensitivity}

\begin{figure}[h] 
\centering
    \includegraphics[width=\columnwidth]{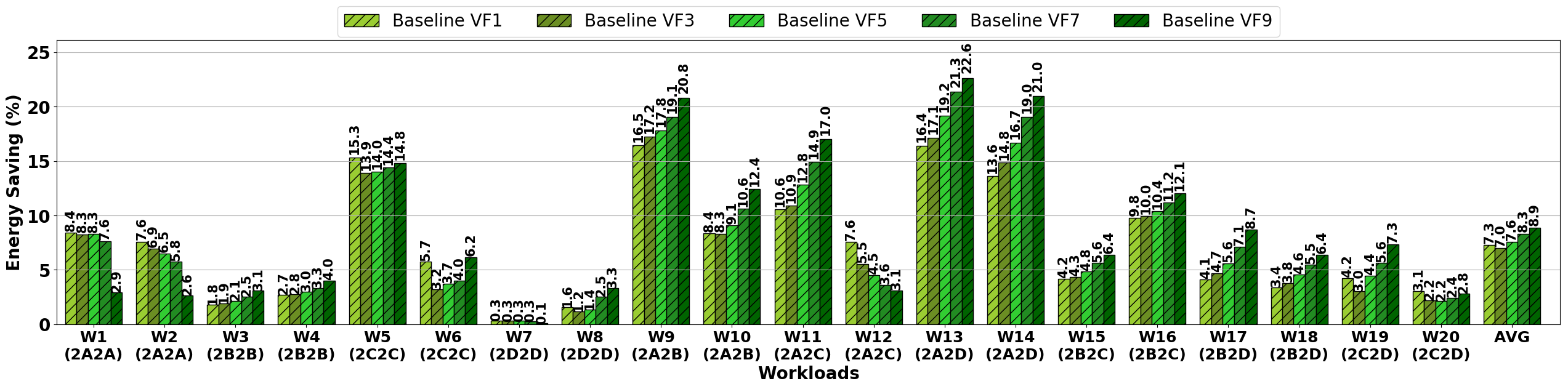}
    \caption{Energy savings for different baseline VF settings using the combined RM.}
    \label{fig_varyingbase}
\end{figure}

Throughout this study, we assumed a mid-range baseline VF. In the last experiment, we evaluate the sensitivity of the proposed \textit{Combined} RMA to different baseline VF settings. This experiment also uses perfect models to exclude the effect of modeling error and QoS violations. Figure~\ref{fig_varyingbase} shows the results of this experiment on 4-core workloads. The bars, from left to right, correspond to baseline VF1, VF3, VF5, VF7, and VF9 out of 10 available VF settings. 

The figure shows two contradicting trends when increasing the baseline VF. This is a result of two opposite effects:
\begin{enumerate}
    \item When the LLC allocation is increased for a cache-sensitive application, it enables a reduction in core VF. This reduction leads to a larger energy saving if the baseline is at higher VF levels, due to the quadratic relation between voltage and energy. 
    \item When selecting a cache-sensitive application as a victim to reduce its LLC allocation, its core VF must be increased. This imposes an energy cost which is higher for larger baseline VF settings. Furthermore, if the baseline is as high as VF9, with only one more higher VF level, it may considerably limit the scope for redistribution of cache resources under performance constraints. 
\end{enumerate}

According to Figure~\ref{fig_varyingbase}, the first effect dominates in most of the cases. This means the proposed RMA is likely to save a higher percentage of system energy with a higher performance target. However, in a few workloads, the second effect dominates which leads to a reduction in percentage of energy savings as the baseline VF increases. This effect is mostly dominant when the victim application is cache sensitive which corresponds to a mix of A or C applications. 

In W1 and W2, a big drop in energy saving occurs when the baseline VF setting increases from VF7 to VF9. In these workloads, victims are selected from type A applications. The VF must be increased by several steps in these applications in order to compensate the performance degradation with additional cache misses. But, there is only one higher VF level available if the baseline is set to VF9. This significantly limits the scope for optimizing the LLC distribution. The second effect also dominates in W12 which is also a mix of cache-sensitive applications. This is, however, not the case in W11. While \textit{bzip2} (type C) in W12 is highly sensitive to LLC allocation with a baseline MPKI around 5 in the dominant phase, \textit{gobmk} (type C) in W11 is very close to the cache sensitivity threshold. Therefore the second effect is not considerable in W11. In some other workloads, including W5, W6, W8, W10, W19, and W20, the second effect dominates only at the lowest baseline VF.

\subsection{Impact of Overheads} \label{subsec_OH_results}

The discussed resource management schemes add overheads to the system in three steps: i) collecting the required statistics, ii) finding the optimal configuration and iii) enforcing the new configuration.

Reading the performance counter values has negligible overhead in an execution interval of 100M instructions. However, the additional instructions that need to be executed for each RMA impose timing and energy overheads. The exact values of these overheads depend on the system configuration at each point of time. Therefore, we evaluate the overhead as a fraction of instruction count to the program execution.

In order to evaluate the instruction count overheads, we implement the proposed RMA as presented in Section \ref{sec_proposed_scheme} in the C programming language. We then compiled and executed this software implementation and measured the number of executed instructions. The number of executed instructions is less than 40K  which is $0.04\%$ of each interval. However, this overhead is accounted for during simulations. 

Finally, when the RMA decides to change the system configuration, there is the overhead of performing DVFS and re-partitioning the LLC. For the DVFS overhead, we assume 15 $\mu s$ and 3 $\mu J$ as reported in \cite{SangyoungPark2013AccurateMicroprocessors} for the Samsung Exynos 4210. The impact of the DVFS overhead is minimal. For example, if the clock frequency is set to 2 GHz and the average IPC is 2, a 100M instruction interval takes 25 ms. In this case, even if the frequency is scaled at every interval, it will add $0.06\%$ to the timing overhead. Both the timing and energy overheads of DVFS are added to the simulation results whenever the RMA chooses a new frequency for each core. Re-partitioning of LLC is limited to modifying a few bit-masks for each core and has negligible overhead.

After re-partitioning, the data movement in LLC happens according to the memory access patterns of applications. The application that receives an additional cache way will gradually replace the data of the previous owner during execution. In our case study, each LLC way contains 256 KB which consists of 4K cache blocks (See Table~\ref{tbl_base_config}). Assuming that all of these blocks will be filled with new data by the new owner --- probably a memory intensive application --- over an interval of 100M instructions, it will cause an additional MPKI of $0.04$ which is negligible compared to the MPKI of memory-intensive applications. Many of these misses may overlap with other misses and do not cause a timing overhead. In reality, a re-configuration happens after several intervals when the program experiences a phase change which further diminishes these overheads.

\section{Related Work} \label{sec_related_work}

Previous attempts to control on-chip resources to enforce QoS constraints on applications include a wide range of types of resources and configuration methods. Adding QoS requirements for the applications has a profound impact on the resource-management approach compared to works that do not take QoS into account \cite{Bitirgen2008CoordinatedApproach,Wang2015XChange:Architectures,Wang2016ReBudget:Reassignment,Jain2017CooperativeNetwork}. A common QoS workload usually consists of a mix of one latency critical (LC) application with strict performance constraints and other best effort (BE) applications \cite{Iyer2007QoSPlatforms,Chen2011PredictiveMultiprocessors,Manikantan2012ProbabilisticPriSM,Funaro2016Ginseng:Allocation,Moreto2009FlexDCP:Architectures,Lo2015Heracles:Scale}. In such cases, the focus is typically to improve the performance of BE applications while providing guaranteed minimum resources for the LC applications. Therefore, the number of LC applications that can run on such a system is very limited and resource optimization is fundamentally dependent of the BE applications. On the other hand, when using DVFS to enforce QoS, energy efficiency can be improved for the LC application \cite{Rahmani2018SPECTR:Management,Choi2002Frame-basedDecoder, Hughes2001SavingApplications,Suh2015DynamicProcessors,Chen2010MemoryOptimization,Pothukuchi2016UsingArchitectures}. However, this prior art does not consider cache partitioning among multiple applications as we do in this paper. Intel Speed-Shift technology \cite{rotem2015intel} is an example of recent DVFS techniques implemented in the Skylake architecture. Compared to the previous Speed-Step technology, which is managed in software, Speed-Shift is managed by the processor, which enables fast and fine-grained control over its voltage-frequency states. Unlike our work, Speed-shift is oblivious to QoS requirements, taking into account only processor utilization. Adding QoS to Speed-shift is an interesting direction to be considered in future work. 

In \cite{Takagi2009CooperativeMultiprocessors,Fu2011Cache-AwareSystems} cache partitioning is used in the proposed solutions, but only to minimize the number of cache misses independently from the DVFS controller. The DVFS controller is responsible for enforcing QoS constraints for workloads, where all applications have QoS constraints. Such an approach is sub-optimal and may potentially lead to QoS violations, since the LLC partitioning controller decides the distribution of LLC allocations without considering its effect on system energy and individual QoS targets. A performance loss due to a reduced LLC allocation must be compensated by an increase in core VF. This may come at a significant energy cost. For some memory intensive applications, it may even be impossible to compensate the performance loss with any of the available VF levels. 

A centralized controller to explore a multi-dimensional design space of different resources is necessary to find the most efficient system configuration. However, the complexity and overhead of such a controller is a serious challenge for online resource management. Many of the previous proposals avoid this issue by breaking the control mechanism into independent controllers for different resources \cite{Lo2015Heracles:Scale, Takagi2009CooperativeMultiprocessors} or different applications \cite{Funaro2016Ginseng:Allocation, Wang2015XChange:Architectures, Wang2016ReBudget:Reassignment}. \cite{Fu2011Cache-AwareSystems} proposes independent controllers for different resources, applications, and even objectives. However, such methods cannot be as efficient as a centralized controller managing several resources because the configuration-space of each local controller is limited under QoS constraints. There have been attempts to come up with coordinated management of multiple resources
 based on machine learning \cite{Bitirgen2008CoordinatedApproach, Jain2017CooperativeNetwork}. The downside of such methods is that they do not provide enough accuracy when applications enter new computational phases. Furthermore, they depend on expensive online learning-processes that are may not be fast enough to react to frequent application phase-changes in multiple concurrently executing applications.

In contrast, in this work, we present a solution to control multiple resources, different objectives, and different applications, in a coordinated fashion, in a centralized controller to maximize the efficiency. We significantly reduce the complexity by intelligently pruning the sections of the design space that lead to inferior results. This method uses statistics from HW performance counters and ATD to model a wide range of resource allocations in a single interval. Such an approach is fast enough to deal with frequent phase changes of applications and provides sufficient accuracy at the new phases. 
\section{Concluding Remarks}\label{sec_conclude}

This paper presents, for the first time, an online Resource Management Algorithm (RMA) that finds the most efficient resource setting, at each program interval, to minimize two important processor energy components, namely core energy-per-instruction and DRAM memory access, using a coordinated controller for DVFS and Last-Level Cache partitioning. It uses simple, yet accurate enough, analytical models to establish the effect of different resource allocations on both performance and energy by collecting statistics from hardware performance counters with no need for any profiling, training or prior knowledge about the detailed run-time behavior of programs. The RMA is implemented in software with appropriate hardware support and invoked at regular program intervals. To keep the run-time overhead negligible, the RMA uses a heuristic algorithm that performs configuration-space exploration in polynomial time.

Our experimental evaluation shows that our combined approach, using coordinated DVFS and cache partitioning, is more effective in saving energy than independent DVFS or cache partitioning RMAs. In addition, the overhead of invoking the RMA at each interval has a negligible impact on the energy savings. The energy savings, when the performance target is the same as the baseline system, can be as high as 18\% and on average 6\%. However, when the QoS target is relaxed to 40\% longer execution time compared to baseline, the proposed RMA can potentially save up to 29\% of system energy and on average 17\%. 

We studied different scenarios where QoS is relaxed only for a subset of the workload. While independent DVFS and cache partitioning RMAs can be more effective when the QoS target is relaxed for applications from specific categories, the energy saving with the proposed combined RMA is higher and less dependent on application categories. We also showed that in the majority of workloads, the  proposed scheme can save a larger percentage of system energy if a higher voltage-frequency level is selected as the baseline which corresponds to a higher performance target. 

\section*{Acknowledgments} 
This research has been funded by the European Research Council, under the MECCA project, contract number ERC-2013-AdG 340328.

\bibliographystyle{elsarticle-num}
\bibliography{references.bib}

\end{document}